\documentclass{llncs}
\usepackage{makeidx}  % allows for indexgeneration
\usepackage{subfigure}
\usepackage{floatrow}
\usepackage[dvipdfmx]{graphicx}

\begin{document}
\frontmatter          % for the preliminaries
\pagestyle{headings}  % switches on printing of running heads
\addtocmark{MEGIITC} % additional mark in the TOC
\mainmatter              % start of the contributions
\title{Results from Pilot Run for MEG II Positron Timing Counter}
\titlerunning{MEGIITC}  % abbreviated title (for running head)
\author{M. Nakao\inst{1} 
\and A. De Bari\inst{2, 3}
\and M. Biasotti\inst{4, 5}
\and G. Boca\inst{2, 3}
\and P. W. Cattaneo\inst{2}
\and \\
M. Francesconi\inst{6, 7}
\and M. De Gerone\inst{4, 5}
\and L. Galli\inst{6}
\and F. Gatti\inst{4, 5}
\and A. Mtchedilishvili\inst{8}
\and \\
D. Nicol\'o\inst{6, 7}
\and M. Nishimura\inst{1}
\and W. Ootani\inst{9}
\and S. Ritt\inst{8}
\and \\
M. Rossella\inst{2}
\and M. Simonetta\inst{2, 3}
\and Y. Uchiyama\inst{9}
\and M. Usami\inst{1}
}
\authorrunning{M. Nakao et al.} % abbreviated author list (for running head)
\tocauthor{M. Nakao et al.}
\institute{Department of Physics, The University of Tokyo, 7-3-1 Hongo, Bunkyo-ku, Tokyo 113-0033, Japan\\
\email{nakao@icepp.s.u-tokyo.ac.jp}
\and INFN Sezione di Pavia, Via A. Bassi 6, 27100 Pavia, Italy
\and Dipartimento di Fisica, Universit\'a degli Studi di Pavia, Via A. Bassi 6, 27100 Pavia, Italy
\and INFN Sezione di Genova, Via Dodecaneso 33, 16146 Genova, Italy
\and Dipartimento di Fisica, Universit\'a degli Studi di Genova, Via Dodecaneso 33, 16146 Genova, Italy
\and INFN Sezione di Pisa, Largo B. Pontecorvo 3, 56127 Pisa, Italy
\and Dipartimento di Fisica, Universit\'a degli Studi di Pisa, Largo B. Pontecorvo 3, 56127 Pisa, Italy
\and Paul Scherrer Institut PSI, 5232 Villigen, Switzerland
\and ICEPP, The University of Tokyo, 7-3-1 Hongo, Bunkyo-ku, Tokyo 113-0033, Japan
}

\maketitle              % typeset the title of the contribution

\begin{abstract}
%The abstract should summarize the contents of the paper
%using at least 70 and at most 150 words. It will be set in 9-point
%font size and be inset 1.0 cm from the right and left margins.
%There will be two blank lines before and after the Abstract.
The MEG II experiment at Paul Scherrer Institut in Switzerland will search for the lepton flavour violating muon decay, $\mu^+\to e^+\gamma$, with a sensitivity of $4\times10^{-14}$ improving the existing limit of an order of magnitude. In 2016, we finished the construction of the MEG II Timing Counter, the subdetector dedicated to the measurement of the positron emission time. The first one-fourth of it was installed in the experimental area and we performed a pilot run with the MEG~II beam of $7\times10^{7}\mu^+/$s. The timing resolution reached the design value improving by a factor of two compared to MEG.
\end{abstract}
\section{Introduction}
The transitions of charged leptons from one generation to another is forbidden in the standard model, therefore it is a background free signature of beyond Standard Model physics.
Actually, many proposed models of beyond  the standard model physics predict charged lepton flavour violating processes at experimentally observable rates.
The best upper limit on the branching ratio for the $\mu^+\to e^+\gamma$ decay was set to $4.2\times10^{-13}$(90\% C.L.) 
by the MEG experiment in 2016\,\cite{MEGfinal}.
We are now upgrading the experiment to MEG~II, using two times higher beam intensity with detectors which have two times improved resolution. In the following the detector dedicated to
the measurement of the positron emission time is described.
% pTC is composed of 512 ultra-fast plastic scintillator with SiPM readouts. The mean hit multiplicity for signal $e^+$ is evaluated to be $\sim$ 9 and a high timing resolution of $\sim$ 35 ps is expected by averaging the signal time of multiple hit counters. To achieve the target resolution, an internal time calibration with a precision of 10 ps or better is required. We have developed two new methods for the calibration, which meet the requirement: Track-based calibration and Laser-based calibration.
%
\section{The Positron Timing Counter}
\begin{figure}[ht]
\begin{tabular}{cc}
\begin{minipage}{0.6\hsize}
	\begin{center}
	\includegraphics[width=\columnwidth]{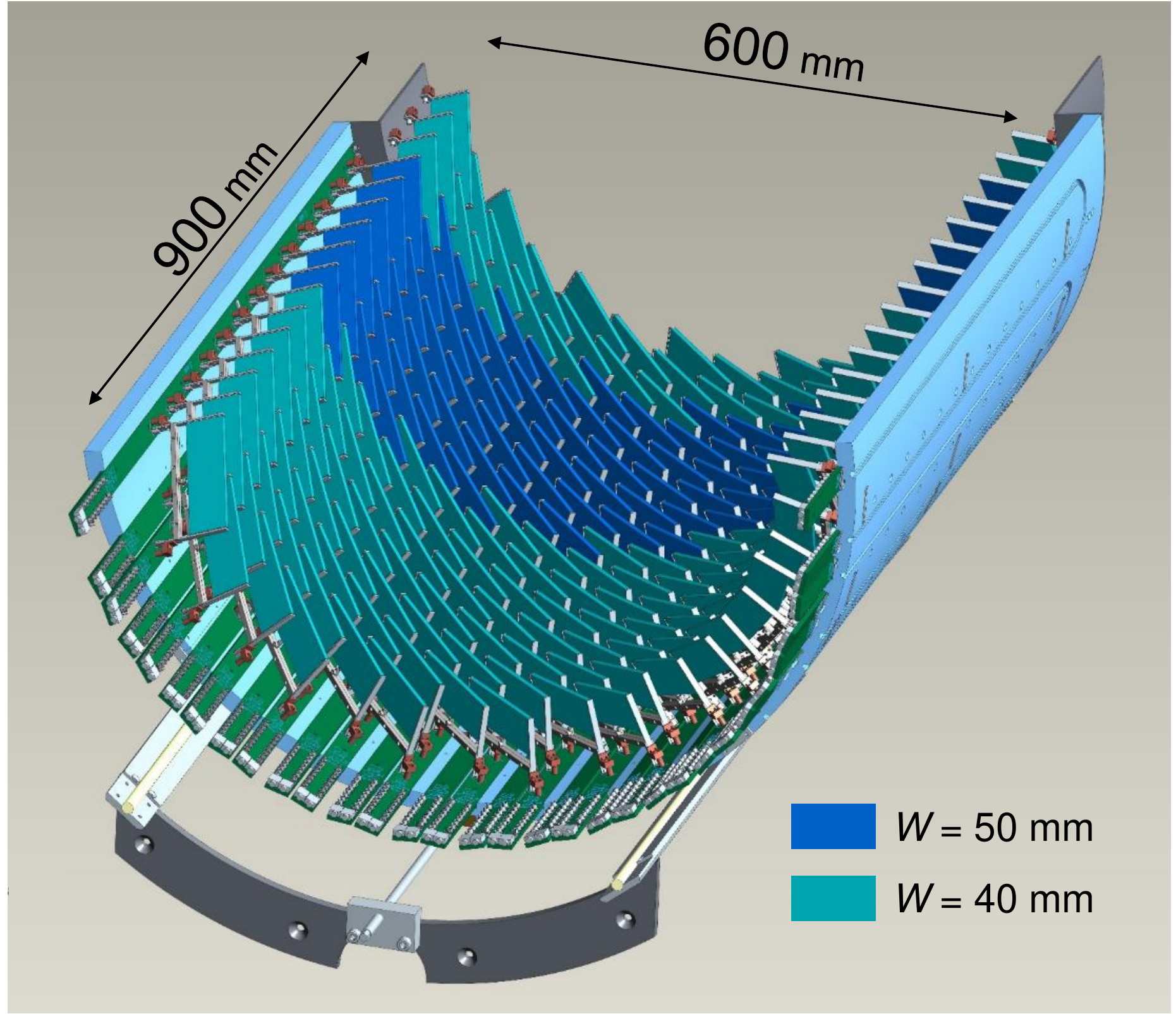}
	\caption{Overview of pTC. Left: Schematic view of downstream pTC. Upstream pTC is mirror symmetric to the downstream about the center of the $\mu$ stopping target.}
	\label{MEG-II-TC}
	\end{center}
\end{minipage}

\begin{minipage}{0.38\hsize}
	\begin{center}
		\subfigure[Picture of counter.]{
		\includegraphics[clip, width=0.8\columnwidth]{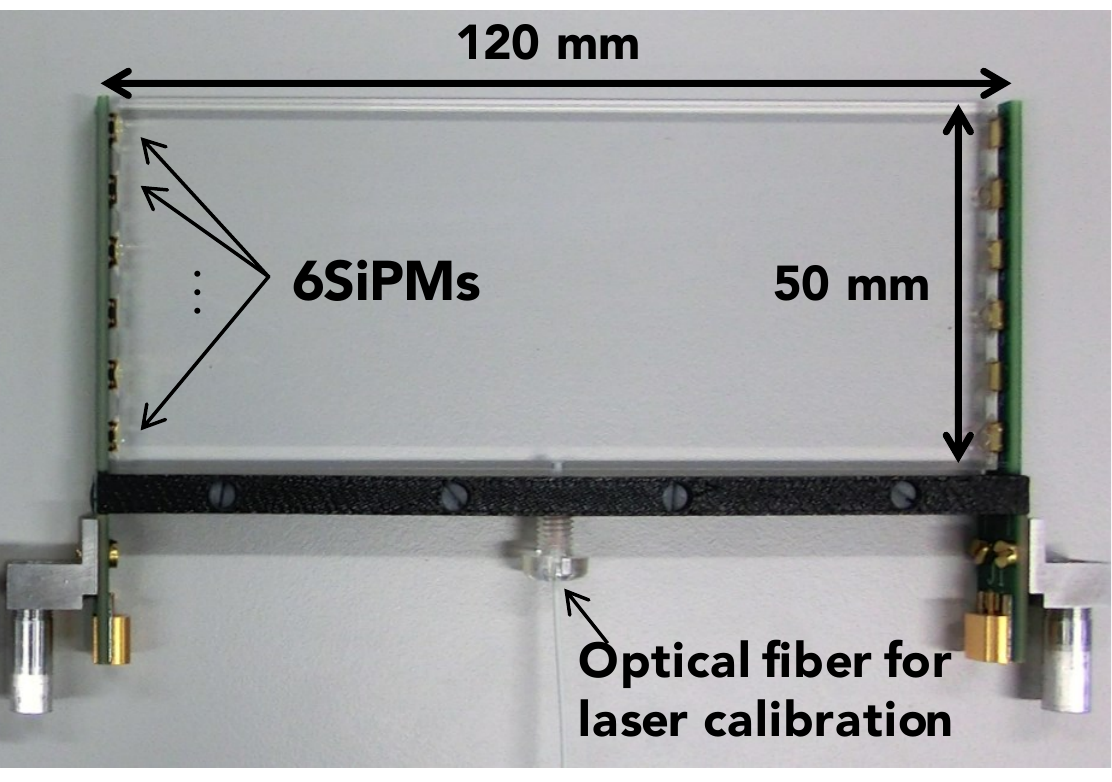}}
		%\label{counter}
		\\
		\subfigure[Probability of hit multiplicity.]{
		\includegraphics[width=\columnwidth]{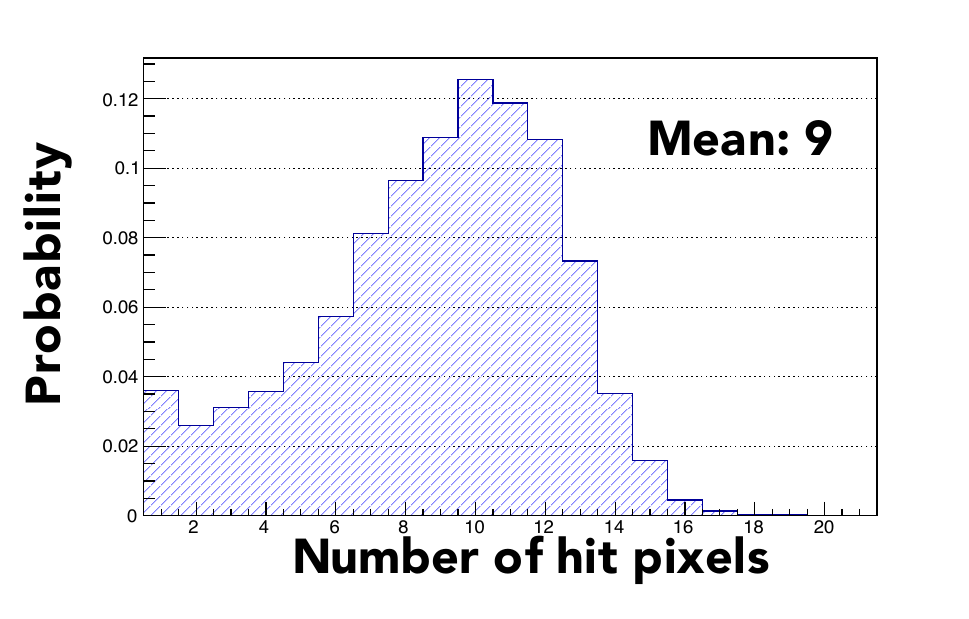}}
		%\label{nhit_prob}
	%\caption{}
	\end{center}
\end{minipage}
\end{tabular}
\end{figure}

\paragraph{Design.}
The pixelated Timing Counter (pTC) is the subdetector in MEG II dedicated to the measurement of the positron ($e^+$) emission time. 
Its design is based on a new approach to improve the $e^+$  timing resolution by a factor of two compared to MEG. 
The pTC is divided into two sectors: upstream and downstream sector, 
each sector consists of 256 scintillation counters as shown in Fig. \ref{MEG-II-TC}. 
The counter shown in Fig. \ref{MEG-II-TC}(a) is made of ultra-fast plastic scintillator\footnote{SAINT--GOBAIN, BC422, 40 (50) $\times$ 120 $\times$ 5 mm$^3$} and read out by six silicon photomultipliers (SiPM)\footnote{AdvanSiD, ASD-NUV3S-P High-Gain, 3 $\times$ 3 mm$^2$, 50 $\times$ 50 $\mu$ m $^2$, V$_{\mathrm{breakdown}}\sim$24\,V} connected in series at both sides.

The key concept of the pTC is to improve the timing resolution by averaging the measured times of multiple hits.
The $e^+$ emitted from the $\mu$ stopping target moves along an almost helical trajectory and hits several counters. 
The averaged hit multiplicity is estimated from MC study as nine as shown in Fig. \ref{MEG-II-TC}(b). 
The total timing resolution is expected to improve with the square root of the number of hits and $\sim$ 35 ps\footnote{$\sim$ 70 ps in MEG} can be achieved with nine hits\footnote{The multiple timing resolution is estimated as $\sigma(N_{\mathrm{hit}})=\sqrt{\frac{\sigma^2_{\mathrm{single}}}{N_{\mathrm{hit}}}+\sigma^2_{\mathrm{const}}}$ where $N_{\mathrm{hit}}$ is the number of hits. 
%1st term is improved with $\frac{1}{N_{\mathrm{hit}}}$ and includes intrinsic timing resolution of each counter and uncertainty among counters which stochastically contributes the total resolution. 
}.

%
%\paragraph{Construction.}
%
\paragraph{Calibration methods.}
Achieving a very precise time resolution with a highly segmented detector requires a fine-tuned time calibration
between the different counters.
We have developed two complementary time calibration methods: the track-based method and the laser-based method. 
The track-based method makes use of $e^+$s from the dominant muon decay (Michel decay, $\mu^+\to e^+\nu\nu$). 
These $e^+$s pass through multiple counters and 
the multiple hits allow to calibrate the time offsets between adjacent counters. 
In addition, a laser pulse is simultaneously divided into each counter for laser-based method. 
This method gives us a direct measurement of time offsets. 
The time calibration accuracy among counters is better than 30 ps.
\section{Pilot Run 2016}
\begin{figure}[ht]
\floatbox[{\capbeside\thisfloatsetup{capbesideposition={left,top},capbesidewidth=6cm}}]{figure}%[\FBwidth]
{\caption{Timing resolution as a function of the number of hit counters ($N_{\mathrm{hit}}$). Points are weighted average of possible counter combinations. The red curve is the best fit function described as $\sigma(N_{\mathrm{hit}})=\sqrt{\frac{\sigma^2_{\mathrm{single}}}{N_{\mathrm{hit}}}+\sigma^2_{\mathrm{const}}}$.}\label{OddEvenResolution}}
{\includegraphics[width=\columnwidth]{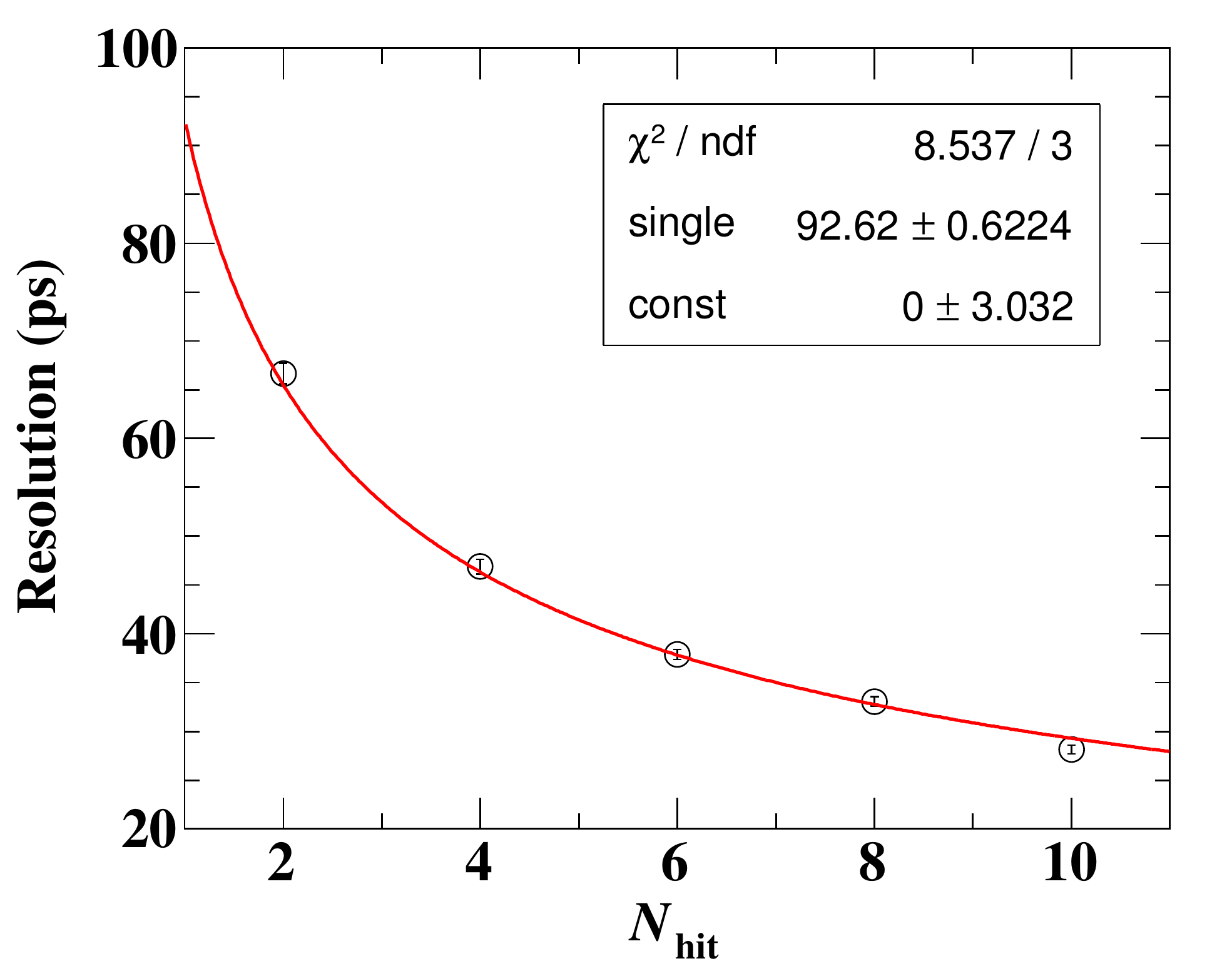}}
\end{figure}

\paragraph{Experimental Setup.}
In 2016, we finished the construction of the pTC and installed the first one-fourth of pTC into the MEG II experimental area to evaluate its performance using the MEG II $\mu^+$ beam ($7\times10^{7}\mu^+/$s) in a pilot run. 
We took data of $e^+$s from Michel decay for 3 weeks.
%and  The overall timing resolution weighted with a distribution of the number of hit counters for signal $e^+$ of 38 ps  was achieved in this pilot run. The prospects towards MEG II physics run are also discussed.
%
An integrated trigger and DAQ system is designed for MEG II and its prototype, called WaveDAQ, was tested during the run. 
%The DAQ system called {\it WaveDREAM} is a multi-functional board which has the functionalities of SiPM biasing, waveform digitization using DRS4 chip\cite{DRS}, amplifier, pole-zero cancellation and first level trigger.
The system is based on the WaveDREAM board, a multi-functional board which has functionalities of SiPM biasing, waveform digitization using DRS4 chip\,\cite{DRS}, amplifier, pole-zero cancellation, and a  part of the trigger processing. Each WaveDREAM board sends pre-processed data to a trigger board which generates the trigger and synchronization signals.
\paragraph{Analysis.}
SiPM output waveforms were sent to the WaveDREAM boards and amplified, shaped, and digitized at a frequency of 2.0 GSPS. Timing information were extracted using the digital constant fraction method \,\cite{dCF}. The fraction was set to 30\%, which gives the best timing performance. The hit time of each counter was calculated by averaging the waveform times at both ends. The timing and position information at counters were used for clustering counters belonging to the same $e^+$ track.
\paragraph{Time calibration.}
We applied both time calibration methods and checked their consistency. The time offsets of each counter calculated independently with the two calibration methods were consistent and stable ($\sim$\,6 ps) during the run. The systematic uncertainty between these methods was 39 ps, which is reduced with $\frac{1}{\sqrt{N_\mathrm{hit}}}$ using multiple hits.
\paragraph{Results.}
The total timing resolution was estimated by an odd--even analysis. First, we selected consecutive counters and divided them into two groups according to their hit order along the track: 
odd ($t_1$, $t_3$, $\cdots$) and even($t_2$, $t_4$, $\cdots$), where $t_{i}$ is the hit time of $i$-th counter. Then, $t(N_{\mathrm{hit}})$ is defined as
t$(N_{\mathrm{hit}})\equiv\left(\sum\limits^{N_{\mathrm{hit}}/2}_{i=1} t_{2i-1}-\sum\limits^{N_{\mathrm{hit}}/2}_{i=1} t_{2i}\right)/N_{\mathrm{hit}}$.
The standard deviation of $t(N_{\mathrm{hit}})$ can be interpreted as the timing resolution in presence of $N_{\mathrm{hit}}$ hits. Fig.~\ref{OddEvenResolution} shows 
that the timing resolution estimated by the odd-even method as a function of $N_{\mathrm{hit}}$. For $N_\mathrm{hit}=9$, $\sigma=31$ ps was achieved from the best fit function. 
The overall timing resolution weighted by the probability of the number of hit counters (Fig. \ref{MEG-II-TC}(b)) was 38 ps. These results correspond to an improvement of 
a factor of 2 compared to the MEG Timing Counter.

\section{Conclusions and Prospects}
We constructed the MEG~II pTC and installed the first quarter. Then, we evaluated the performance under the MEG II environment. The overall timing resolution of 38 ps was achieved as required. We are currently preparing for another pilot run in the end of 2017. We will install other detectors such as Liquid Xenon Gamma-ray Detector and Radiative Decay Counter in addition to the full pTC. 
Six weeks of combined detector data taking are planned under the MEG II environment. The full engineering run will start in 2018 followed by the physics run. The sensitivity of $4\times10^{-14}$,
improving the existing limit of an order of magnitude, can be achieved in three years' data taking.

\clearpage
\addtocmark[2]{Author Index} % additional numbered TOC entry
\renewcommand{\indexname}{Author Index}
\printindex
\clearpage
\end{document}